\documentclass[apj]{emulateapj}

\shorttitle{Spatially Resolved Spectroscopy of Elias~1}
\shortauthors{Goto et al.} 

\begin{document}

\title{Spatially Resolved 3~\micron~Spectroscopy of Elias~1:\\
Origin of Diamonds in Protoplanetary Disks \altaffilmark{1}}

\author{M. Goto,\altaffilmark{2}
Th. Henning,\altaffilmark{2}
A. Kouchi,\altaffilmark{3}
H. Takami,\altaffilmark{4}
Y. Hayano,\altaffilmark{4}
T. Usuda,\altaffilmark{4}
N. Takato,\altaffilmark{4}
H. Terada,\altaffilmark{4}
S. Oya,\altaffilmark{4}
C. J\"ager,\altaffilmark{5}
A. C. Andersen\altaffilmark{6}
}

\email{mgoto@mpia.de}

\altaffiltext{1}{Based on data collected at Subaru Telescope, which is
                 operated by the National Astronomical Observatory of
                 Japan.}

\altaffiltext{2}{Max Planck Institute for Astronomy, 
  K\"onigstuhl 17, D-69117 Heidelberg, Germany.}

\altaffiltext{3}{Institute of Low Temperature Science, Hokkaido
University, Sapporo 060-0819, Japan.}

\altaffiltext{4}{Subaru Telescope, 650 North A`ohoku Place, Hilo,
                 HI 96720, USA.}

\altaffiltext{5}{Institut f\"ur Festk\"orperphysik, 
   Friedrich-Schiller-Universit\"at Jena, Helmholtzweg 3, D-07743 Jena,
   Germany.}

\altaffiltext{6}{Dark Cosmology Centre, Niels Bohr Institute, University
of Copenhagen, Juliane Maries Vej 30, DK-2100 Copenhagen, Denmark.}

\begin{abstract}

We present spatially resolved 3~$\mu$m spectra of Elias~1 obtained with
an adaptive optics system. The central part of the disk is almost devoid
of PAH emission at 3.3~$\mu$m; it shows up only at 30~AU and beyond. The
PAH emission extends up to 100~AU, at least to the outer boundary of our
observation. The diamond emission, in contrast, is more centrally
concentrated, with the column density peaked around 30~AU from the
star. There are only three Herbig Ae/Be stars known to date that show
diamond emission at 3.53~$\mu$m. Two of them have low-mass companions
likely responsible for the large X-ray flares observed toward the Herbig
Ae/Be stars. We speculate on the origin of diamonds in circumstellar
disks in terms of the graphitic material being transformed into diamond
under the irradiation of highly energetic particles.

\end{abstract}


\section{Introduction}

Spatially extended PAH emission, but confined to a protoplanetary disk
\citep{hab04b,boe04,gee07,dou07}, provides crucial evidence for the
two-layer disk model consisting of a cold mid-plane and a warm disk
atmosphere \citep{men97,chi97,dul01}. More detailed studies on PAH
chemistry coupled with advanced disk modeling may provide a powerful
tool for using PAH emission to better understand the evolution and
structure of protoplanetary disks \citep{vis07,dul07}. However, the
emission feature at 3.53~$\mu$m observed in the spectra of several
Herbig Ae/Be stars has been an enigma since its discovery in the 1980s
\citep{whi83,whi84}. Many explanations have been proposed
\citep{sch90,all92}, until \citet{gui99} conclusively identified small
diamond particles as the carrier of the emission
\citep{she02,jon04}. Despite extensive investigation
\citep[e.g.,][]{ker02,hab04a,top06}, the origin of diamonds in
protoplanetary disks remains essentially inconclusive, except that they
are likely not of interstellar origin but are formed close to the star.

Elias~1 (V~892~Tau) is a Herbig Ae/Be star located in the Taurus-Auriga
complex at a distance of 140$\pm$20~pc \citep{eli78}, with an estimated
age younger than 3~Myr \citep{str94}. The geometry of the disk is poorly
known. A large infrared excess is still present in the far-infrared
spectral energy distribution, indicating flaring in the outer part of
the disk \citep{mee01}. Near-infrared speckle observations detected a
possible elongation of 40 to 100~AU from east to west
\citep{kat91}. However, whether it represents a disk or a scattered
bipolar lobe is still an open question \citep{haa97}. The source is
unresolved at sub-mm wavelengths, leaving little useful information as
to the inclination or the disk position angle
\citep{hen98,dif97}. However, recent mid-infrared nulling interferometry
observations give a preliminary disk diameter of 20--30 AU at 10~$\mu$m
\citep{liu07}.

\section{Observations} 

Long-slit spectroscopy was performed at the Subaru Telescope on
September 12, 2003, using the facility spectrograph IRCS
\citep{tok98,kob00} together with a curvature-sensing adaptive optics
system optimized for nearly diffraction-limited imaging at wavelengths
longer than 2~$\mu$m \citep{tak04}. The medium-resolution grism was used
with a reflective slit of 0\farcs3 $\times$ 20\arcsec~to provide spectra
with a resolving power of $R =$ 1,000. The slit was oriented in two
directions, one east-to-west parallel to the major axis of the
elongation of the disk suggested by \citet{kat91}, and one perpendicular
to this direction. The spectra were recorded by moving the tip-tilt
mirror inside the adaptive optics system by 3\arcsec~along the slit for
sky subtraction. The amplitude of dithering virtually defined the outer
boundary of our observation. Extra caution was taken to center the
object in the slit in order to prevent recording the point spread
function (PSF) in its wings. The short integration time of 2~s was
repeated 10~times for one exposure to avoid detector saturation. The
spectra of the standard star HR~1490 ($V = 5.8$~mag; A2V) were obtained
immediately after Elias~1, also with the adaptive optics system. The
standard star was dimmed by the neutral density filters in front of the
wavefront sensor so that the residual wavefront error was as similar as
possible to that of Elias~1 ($V=$15.3~mag).

\section{Data Reduction and Results}

\subsection{Spectral extraction}

\begin{figure}
\begin{center}
\includegraphics[width=0.56\textheight,angle=-90]{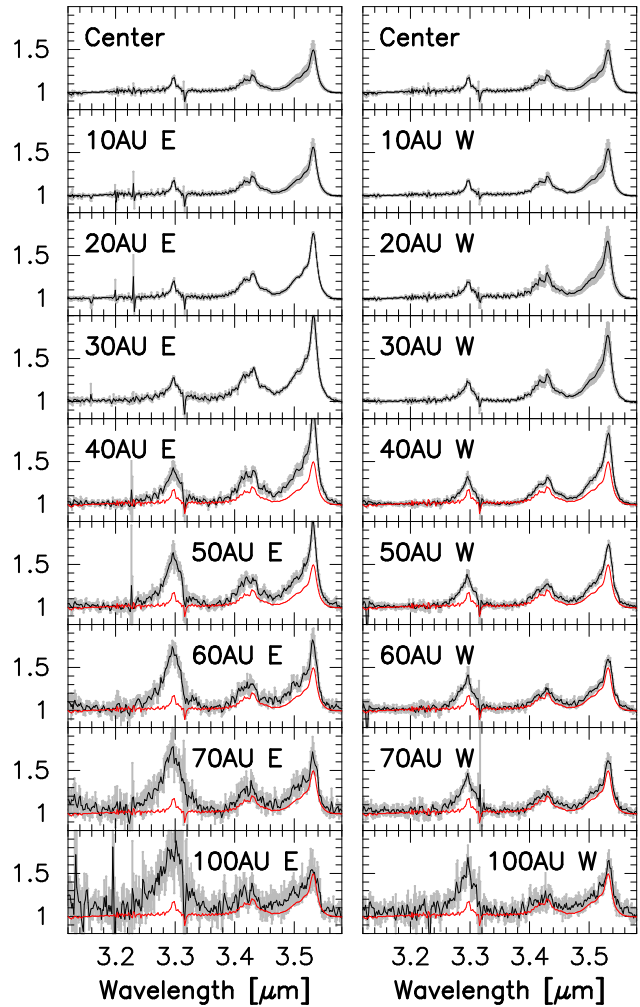}
\end{center}
\caption{A spectral sequence across Elias~1 from east to west. The
spectra were extracted every 10~AU from the star up to 70~AU, and at
100~AU (80--120~AU) at the outer boundary of the observation.  The
actual instrumental resolution is likely 20--30~AU. The red curves
duplicate the spectra from the center to emphasize the variation. The
error bars of the data points are given in gray. All spectra were
normalized by the polynomial function, fitted to the continuum
emission. The relative intensity of PAH emission (3.3~$\mu$m) to
diamond emission (3.53~$\mu$m) becomes larger with distance, in
particular in the outer region beyond 30~AU from the star. The
pointed emission at 3.3~$\mu$m in the central spectrum, much narrower
than the PAH feature, is Pf~$\delta$. The emission at 3.42~$\mu$m is
also from diamond, but from its defect. \label{f1}}
\end{figure}

\begin{figure}
\begin{center}
\includegraphics[width=0.56\textheight,angle=-90]{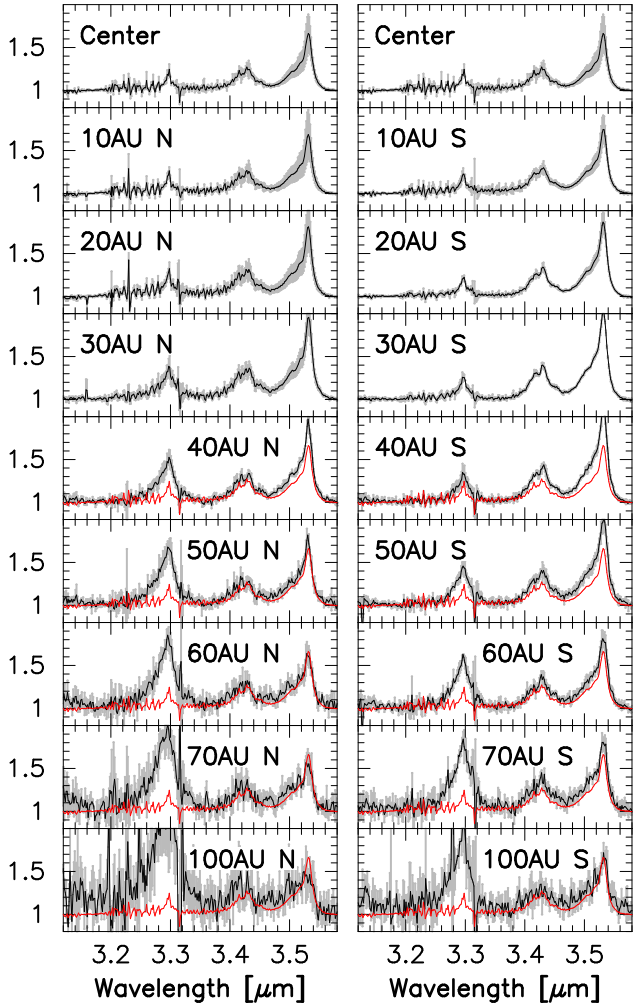}
\end{center}
\caption{Same as Figure~\ref{f1}, but a spectral sequence along the
north to the south. The same trend of the variation of PAH emission
relative to the diamond is seen here. \label{f2}}
\end{figure}

Two-dimensional images of the spectra were reduced separately frame by
frame in order not to smear out the spatial resolution. After
preliminary processing, including pair subtraction and flat fielding,
the centroid of the PSF was traced along the wavelength to set the
aperture center of the spectral extraction. In total, 17 strips of
spectra were extracted from one frame from the apertures set at every
10~AU (0\farcs07) from the center up to 70~AU (0\farcs5) on both sides
of the star, and from one additional aperture at 100~AU (80--120~AU) at
the outer edge. The aperture width (0\farcs07) was set for a proper
sampling of the PSF (FWHM$\approx$0\farcs1). However, it was about the
same size as the pixel scale (0\farcs06). The limited sampling of the
PSF, together with a slight curvature of the spectra along the array
column, sometimes causes an artificial wobbling of the spectral
continuum. In order to minimize this effect, we extracted the same
number of spectra from the standard star HR~1490 across its PSF and
divided the spectrum of Elias~1 by the spectrum of HR~1490, extracted
from the same aperture with the same offset from the center. The
wavelength calibration was performed by matching the telluric absorption
lines to the model transmission curve computed by ATRAN \citep{lor92}. A
set of spectra extracted from different frames, but at the same aperture
were finally combined, and the dispersion of the spectra was taken as
the error of the observation (Figure~\ref{f1} and ~\ref{f2}).

From the data, it is obvious that the PAH emission is relatively weak in
the center (note that the pointed feature at the center at 3.3~$\mu$m is
from Pf~$\delta$). The trend is identical for the east-to-west and the
north-to-south directions. The spatial variation at a physical scale of
about 10~AU already indicates that the PAH emission indeed arises from
the circumstellar disk and not from a large envelope.

\begin{figure}
\begin{center}
\includegraphics[width=0.48\textwidth]{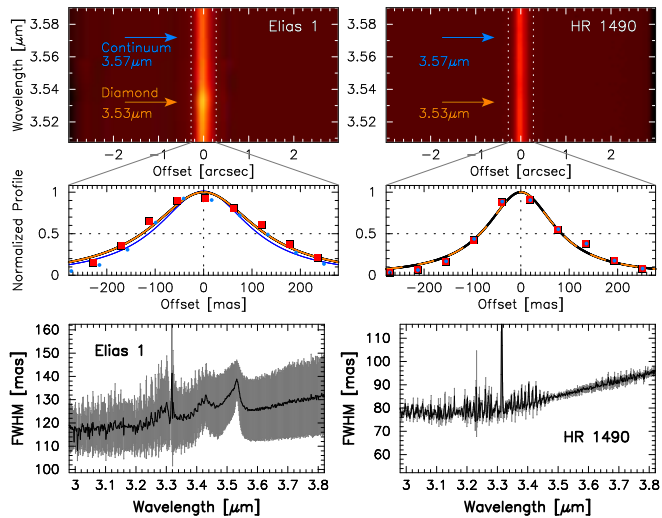}
\end{center}
\caption{The spatial profile of the diamond emission. Top left and
middle left: The diamond emission (3.53~$\mu$m) is slightly extended
compared to the continuum emission (3.57~$\mu$m) right next to the
emission features. Bottom left: The FWHM of the emission profiles
measured along the wavelength. There is unambiguous ehancement of the
FWHM at the PAH and the diamond features. Right: Same as the left
panels, but for the standard star HR~1490. There is no trace of extended
emission. If we take the FWHM of the nearby continuum (126~mas) to
deconvolve the diamond profile (139~mas), the physical extension is
8~AU. If we instead take the FWHM of the standard star at the same
wavelength (86~mas), it amounts to 15~AU. Note that the numbers
represent the minimum extension of the diamond emission in both cases,
where the emission feature contributes a substantial fraction of the
flux at the wavelength comparable to the continuum emission.\label{f3}}
\end{figure}

The spatial extension of the diamond emission is less obvious. We double
checked the spatial profile of the diamond emission with respect to the
continuum emission (Figure~\ref{f3}). The FWHM of the monochromatic PSF
at the diamond feature is apparently larger than the one at the
neighboring continuum emission. If we deconvolve the diamond emission
with the continuum PSF, we could set the minimum size of the diamond
emitting zone to 8--15~AU.

\subsection{Surface density}

The absolute line flux was calibrated with respect to the standard star
HR~1490, as Elias~1 itself is apparently extended, and the loss at the
slit transmission is uncertain. For the calculation of the surface
density of the PAH molecules, we assumed that all the energy absorbed in
the UV to visible wavelengths is re-emitted in the infrared. Because we
are more interested in the relative spatial variation, only a single
size of PAHs was used in the calculation for simplicity. The absorption
cross-section of a neutral PAH ($a$=5.02$\times 10^{-4}~\mu$m;
$N_C\approx$ 58) was taken from \citet{li01}.

The spectral type of Elias~1 is uncertain, depending on how much
foreground extinction is assumed, and varies from A6 \citep[e.g.,
][]{coh79} to B9 \citep[][]{str94}. Here we take A6e from \citet{the94}
and use the photospheric model of A6V \citep{kur79} with the luminosity
scaled to $L_\ast = 56~L_\odot$ \citep{ber92}. No extinction of stellar
radiation was applied along the line of sight. PAH molecules of this
size are small enough to be heated stochastically. We refer to
\citet{dl01} for the conversion of internal energy to the emitting
temperature of the PAH molecules.

\begin{figure}
\begin{center}
\includegraphics[width=0.45\textwidth,angle=-90]{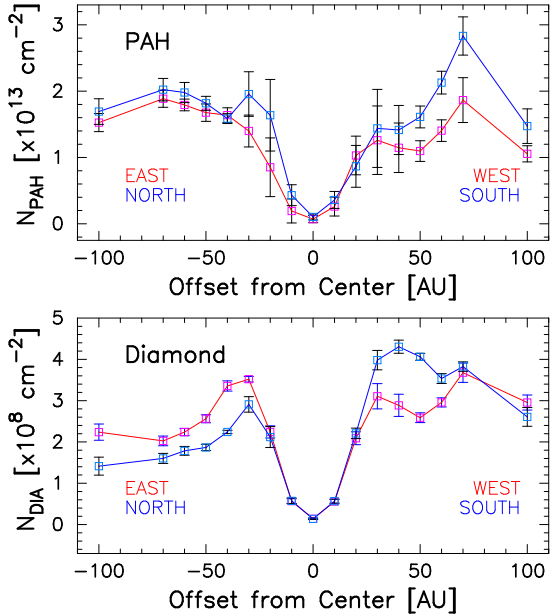}
\end{center}
\caption{The radial profiles of the surface density of PAH and
diamonds. \label{f4}} 
\end{figure}

The same calculation was repeated for diamonds, although the size of the
diamond particles is still debated. The hydrogenation and the surface
lattice structure require the temperature of diamond to be between 800
and 1000~K. \citet{ker02} argued that the particle has to be as small as
1--10~nm to consort with the range, as diamonds are poor emitter in the
infrared, and larger particles become too hot and decompose
themselves. However, the sheer appearance of the 3.53~$\mu$m feature
requires well-ordered crystal lattice of the physical scale of at least
25--50~nm \citep{she02}, or even larger \citep{jon04}. Here we assume a
size of 100~nm, with the diamond particles being in equilibrium with the
radiation. The absorption cross-section in the UV to the visible
wavelengths is taken from \citet{mut04} and combined with that of
hydrogenated diamond, which is responsible for the 3.53~$\mu$m emission
\citep{che97}. The spatial variation of the surface density of the PAH
molecules and diamond particles is shown in Figure~\ref{f4}.

PAH molecules are much less abundant close to the star, within about
30~AU from the center. The abundance becomes higher with radial distance
up to the outer boundary of the observation. However, the diamond
particles show a distinct spatial distribution from the PAHs and are
more centrally concentrated but with peaks near 30~AU, where the PAH
emission starts to emerge.

\section{Discussion}

\subsection{How Special Is a Diamond Emission Star?}

There are only 3 Herbig~Ae/Be stars known that have clear diamond
signatures at 3.53~$\mu$m: HD~97048 \citep{whi83}, MWC~297
\citep{ter01}, and Elias~1 \citep{whi84}. An extensive survey conducted
by \citet{ack06}, covering more than 60 Herbig~Ae/Be stars with 3~$\mu$m
spectroscopy, did not add a new source to the ones already
known. However, the survey did find a few additional sources that might
show the diamond features, such as T~CrA, V~921~Sco, and
HD~163296. There have been several reports of the detection of emission
near 3.5~$\mu$m, for instance HD~142527 by \citet{wae96} and HD~100546
by \citet{mal98}. However, these newly found, or suggested, sources show
relatively weak emission at 3.5~$\mu$m, if any at all, but nothing
comparable to those seen in the three Herbig~Ae/Be stars listed
above. The original paper of \citet{whi83} that first reported the
3.53~$\mu$m emission from Elias~1 and HD~97048 included TY~CrA as
another possible star that might show the same signature. However, the
emission was not reproduced in later observations at the same wavelength
\citep{ack06}. In contrast, the presence of diamond emission in Elias~1
and the other two Herbig~Ae/Be stars has been demonstrated by multiple
observations. We therefore take only these three sources as genuine
detections. We discuss below if and how they are special and why they
are so rare.

There are only two Herbig Ae/Be stars known thus far that have shown
large X-ray flares: Elias~1 and MWC~297 \citep[][]{gia04,ham00}. Aside
from the coincidence, this is intriguing, because Herbig Ae/Be stars are
not supposed to be X-ray sources themselves. OB stars and young low-mass
stars are soft and hard X-ray sources, respectively, due to the strong
stellar winds and their magnetic activity \citep[for a review
][]{fei07}. In contrast, Herbig Ae/Be stars are too hot to maintain
convective zones at the surface but are not luminous enough to set off
strong stellar winds.

However, the X-ray flares from Elias~1 and MWC~297 have been reasonably
accounted for by their low-mass companions. In addition to a T~Tauri
star at 4\arcsec~away \citep[V~892~Tau~NE; ][]{gia04}, Elias~1 was
further resolved into two sources by bispectrum speckle interferometry
\citep{smi05}. The newly found companion, only 50~mas or 7~AU away from
Elias~1, is expected to have a mass in the range of 1.5 to 2~$M_\odot$,
very close, but within the upper limit of stellar mass that can still
have a convective outer layer. Likewise, more than a handful X-ray
sources cluster around MWC~297 \citep{dam06}. One of the faint sources
detected at $H$ with an adaptive optics system at 3\farcs4 away from
MWC~297 is comparable to a T~Tauri star in its brightness
\citep{vin05}. The plasma temperature going up to $\sim$ 8~keV during
the flare seen in Elias~1 \citep{gia04} indicates that it is caused by a
similar mechanism for a solar flare. The unusual X-ray flares toward the
Herbig Ae/Be stars are most likely attributable to the low-mass
companions found recently \citep{ste06}.

HD 97048, the last Herbig Ae/Be star with diamond emission, is also
known to be an X-ray source \citep{zin94}. No temporal variability of
X-ray emission is reported. The search for very close companions is so
far negative \citep[e.g.,][]{ghe97}. However, HD~97048 is among the
hardest X-ray sources in Cha~I cluster studied by XMM~{\it Newton}
\citep{ste04}. Along with the presence of nearby low-mass X-ray sources,
it may imply that the circumstances are similar to those of Elias~1 and
MWC~297.

Yet another example of a diamond signature is a post-AGB star, HR~4049
\citep{geb89}. Although the evolutionary stage is totally different, it
is tempting to note that the star's spectral type is similar to that of
the Herbig Ae/Be stars (B9.5~Ib-II); it also has a disk around
\citep{dom03} and even a possible white dwarf companion as well
\citep{lug05}.

A disk, a hot central star, and a companion emitting hard X-rays seem to
be the conditions that ``diamond stars'' must have. The particular suite
of conditions already suggests that the circumstellar environment plays
a critical role in the formation of diamond, or at least the diamond
features to appear. However, the nano-diamond particles found in
meteorites are widely considered to be presolar origin, because of the
isotope anomaly incompatible with that of solar system \citep{lew87}.

Searching sources of diamond outside the solar system is greatly in debt
to the laboratory study of diamond formation facilitated by the interest
from material science. \citet{tie87} drew parallel between the synthesis
of detonation diamond \citep[e.g.][]{bai07} with possible diamond
formation in the grain-grain collisions in the strong shocks in the
ISM. \citet{kou05} used experimental simulation of the icy grains in the
diffuse ISM to demonstrate diamond cores of nanometer size grow in the
organic refractory in the UV irradiation.

Chemical vapor deposition (CVD) of diamond has been most extensively
studied as the low-pressure path of diamond growth in the ISM. The CVD
technique has been used to grow diamonds on the substrate in warm
environment ($\sim$ 1000~K) from hydrocarbon gas dissociated either by
hot filaments or in a microwave cavity while protecting the growing
surface in the hydrogen-rich atmosphere \citep[for a
review][]{ang88,fre89b}. Homogeneous nucleation relevant to the diamond
formation in the ISM is also demonstrated in the laboratory by
\citet{fre89a}, although with chloride which is alien to the ISM but
plays critical role in the enhancement of the nucleation rate
\citep{asm99}. As the laboratory conditions of CVD experiments is
similar to the physical conditions of atmosphere of carbon stars, much
effort has been gone into linking the CVD diamond to astronomical
observations \citep[e.g.][]{wdo87,and98,she02} and to the nano-diamonds
extracted from carbonaceous meteorites \citep{dau96}. CVD diamond also
inspired possible formation of diamond in supernova outbursts
\citep{cla95,nut92}. \citet{jor88} combined the two sources together,
and tried to explain the isotope anomaly by a single binary system
consisting of a carbon star and a white dwarf that eventually undergoes
supernova outburst.


However, the laboratory conditions of CVD formation calls for an extreme
carbon rich atmosphere in the standard of the ISM. A carbon star is a
rare exception where the carbon abundance exceeds that of oxygen, and
hydrocarbon molecules are copious in the environment. On the contrary,
the ISM is intrinsically oxygen rich, so as a protoplanetary disk
is. The abundance of carbon that \citet{wdo87} refers is $n({\rm C}) = 4
\times 10^{13}$~cm$^{-3}$ in the expanding atmosphere of the pressure
10$^{3}$~dyn~cm$^{-2}$ where $n({\rm H)}$/$n({\rm C}) =$200. On the
other hand, the density of a protoplanetary disk is $n({\rm
H})=10^{11}$~cm$^{-3}$ in the innermost region where the conditions are
most favorable \citep{aik06,jon07}. The carbon density including all
atoms and molecules combined together is $n({\rm C})=10^{7}$~cm$^{-3}$
assuming nominal fractional abundance of carbon in the ISM. If we scale
the growth rate of diamond in CVD formation on a warm substrate
80~$\mu$m~hr$^{-1}$ \citep{ang88} to the carbon abundance of a
protoplanetary disk to a carbon star, a diamond particle in a
protoplanetary disk could possibly grow to 200~nm in 1~Myr, which is
about the size that we expect the diamond particles are in
Elias~1. However, this is overly an optimistic estimate where no carbon
atoms are locked in carbon monoxide, and all available for the diamond
formation. In reality, the most abundant hydrocarbon C$_2$H in a
protoplanetary disk is expected to be an order of 6--8 less than the
total carbon abundance [$n({\rm C_2H}) \approx 10^{-1}$~cm$^{-3}$].
Moreover, the growth rate implied by the homogeneous nucleation by
\citet{fre89a} is 10~nm~hr$^{-1}$, which is 3 orders of magnitude lower
than we assumed above.


Nevertheless, in contrast to PAHs and silicates commonly observed both
in disks and in the ISM, no diamonds emission or absorption has been
ever detected in the diffuse ISM, which strongly implies that the
diamonds in Elias~1 are formed {\it in situ}. \citet{dai02} argued that
the depletion of diamond particles in certain types of interplanetary
dust is evidence that some diamonds are formed in the inner solar
nebula. Even though the diamonds in the solar system is indeed pre-solar
origin, that causes no contradiction with the diamond in Elias~1 being
formed in the protoplanetary disk, as they may not have a common origin
in the first place. The diamond particles extracted from meteorites are
relatively small, between 1--10~nm in diameter \citep[e.g., ][]{ber90},
while the particle size of diamonds in Elias~1 and HD~97048 is at least
one order of magnitude larger.  Moreover, only 3 Herbig Ae/Be stars
among more than 60 investigated show clear diamond emission
features. There is no way to argue that diamond formation in the disks
of Elias~1 and other Herbig Ae/Be stars is something other than an
exception.





\subsection{Diamonds in the ISM and Protoplanetary Disks}

\subsection{The Carbon Onion as a Pressure Cell}

We propose here a diamond formation route in a protoplanetary disk using
carbon onions as the high-pressure cell. The obvious problem in forming
diamonds in a protoplanetary disk is how to maintain high pressures
\citep[$>$1~GPa;][]{bun89}, because graphite is otherwise energetically
more favorable when the particles are macroscopic size. \citet{ban96}
and \citet{ban97} produced carbon onions from nanometer-sized graphite
particles by the application of electron beams (1.25~MeV; 200
A~cm$^{-2}$). The carbon onions made in this way have concentric shells
but with a spacing significantly smaller (0.31~nm) than in standard
graphite material (0.34~nm). This is because the carbon atoms in the
outer shells are knocked on during electron irradiation, and the
outermost shells consequently shrink inward. The spacing of the shell
becomes smaller in the inner part up to 0.22~nm at the center. The
nucleus of the carbon onions therefore experiences enormous pressure
estimated to be 50 to 100~GPa. The multiple onion-like shell therefore
virtually provides a high-pressure cell required for diamond
formation. \citet{ban96} further applied high-energy electron beams to
the carbon onions and demonstrated that the cores of the carbon onions
are indeed transformed into the lattice of the diamond within an hour or
even faster.

It is not easy to show direct evidence of large carbon onions by
astronomical observations because they should rather have a continuous
absorption without bands or lines due to functional groups. There are no
absorption measurements of such particles in laboratory because the
perfect onions are only stable in the microscope under heating.
Nevertheless, there are indications of carbon onions in the ISM from
theory, laboratory experiments, and astronomical observations, though
their sizes are not necessarily large. \citet{tom93} has shown that when
the size of the carbon cluster is more than 20 carbon atoms, a spherical
multi-shell cluster is energetically most favorable among conic or
cylindrical shells, or a single large shell. Laboratory experiments with
resistively heated graphite rods by \citet{sch98} and \citet{jae99} have
shown that the presence of hydrogen atoms like those in the ISM
introduces curvature in the graphene planes, leading to the formation of
an onion-like structure. The interstellar extinction curve shows a bump
at 217.5 um with a remarkably stable central wavelength and narrow bump
width. This might be also explained by carbon onions in the ISM
according to the model calculation by \citet{hen97} and
\citet{tom04}. Carbon onions have been found in meteorites
\citep[e.g. in Murchison by ][]{ama93}. Although most of them are much
smaller than 100~nm, the carbon onion particles found in Allende
meteorite are as large as 10 to 50~nm \citep{smi81}.

Although we have only circumstantial evidence of large carbon onions in
the ISM, they could also be formed in the same way as diamond under the
electron or ion irradiation. Note that \citet{ban96} experiments do not
start with carbon onions but are prepared beforehand from small graphite
particles by electron irradiation. The transformation of carbon soot to
multi-shell carbon particles (about 50 nm in diameter) under electron
irradiation was also experimentally demonstrated by \citet{uga92}.

\subsection{Surfacing Diamond}

The emission features at 3.5~$\mu$m come from vibrational stretching of
$sp^3$ C-H bonds on the surface of diamond particles. Even if the
diamond is formed in the core of the carbon onion pressure cell, the
diamond emission features are not visible as long as it is shielded by
the onion shells. The diamond has to surface and be covered by ambient
hydrogen atoms if the spectral features are to be observed.

The reversal of phase stability in diamond-graphite systems induced by
irradiation offers a natural process for surfacing the diamonds.
\citet{zai97} and \citet{zai00} have shown that once the diamond
nucleates in the core of a carbon onion, the formation of diamond
continues to the surface of the onion at the expense of the graphite
shells until the entire carbon onion turns into a diamond particle.
Graphite is more subject to displacement under the nonequilibrium state
induced by constant particle irradiation, because of its lower threshold
energy ($\approx$10--20~eV) than that of diamond \citep[$\approx$ 35~eV;
][]{zai97}. In an extreme circumstance where the carbon atoms are
disturbed by uninterrupted particle impacts, the graphite shells are
therefore gradually converted into diamond, which is the more stable
form. 

The most interesting aspect of the irradiation-induced phase
transformation is that the displacement cross-section of graphite with
respect to that of diamonds peaks at a temperature of $\sim$600~K
\citep{zai00}. At that temperature the transition from graphite to
diamonds does not require high pressure but proceeds as long as diamond
nuclei exist \citep{lyu99}.

Low-mass stars during active accretion are hard X-ray sources, creating
energetic particles required for diamond formation.  However, diamond
particles are only seen in warm disks around Herbig Ae/Be stars; none
are found in T~Tauri stars. This might be most simply understood in the
same way as why fewer T~Tauri stars show PAH emission than Herbig Ae/Be
stars. Out of 54 pre-main sequence stars studied with {\it Spitzer}/IRS
by \citet{gee06}, 5 out of 9 Herbig Ae/Be stars show PAH emission,
whereas only 3 out of 38 T~Tauri stars do so. There might be plenty of
diamonds in T~Tauri disks, but we do not see them at all. This is
because of insufficient radiation either to keep diamond particles in
the warm disk atmosphere or to excite the C-H vibrational transition, as
the effective temperatures of T~Tauri stars ($T_{\rm eff} < 6000$~K) are
lower than those of Herbig Ae/Be stars ($T_{\rm eff} \sim 10000$~K).

Another attractive explanation would be that the surfacing of diamond is
hindered by the relatively cool disks of T~Tauri stars where the stellar
luminosity ($L_\ast =0.1$--$25~L_\odot$) is orders of magnitude less
than that of Herbig Ae/Be stars. The surface area of the warm region in
the disk ($>600$~K) where diamond formation proceeds without high
pressure becomes smaller by a factor of 2--500 in T~Tauri stars. The
diamonds may exist in the grains in the disks of T~Tauri stars, but they
escape observation because they are covered under the mantles of carbon
onions without C-H bonds on the surface to produce the diamond emission
at 3.53 $\mu$m. This might have consequence why the diamond particles
sampled in the meteorites in the solar system are all small in the
nanometer size \citep{lew87}.


\subsection{Origin of diamond}

We present below a summary of diamond formation in the disk of
Elias~1. The X-ray flares seen toward Elias~1 and MWC~297 originate from
the magnetic activity of low-mass companions. This is a signpost of the
acceleration of high-energy particles. The presence of PAH emission, in
contrast, suggests the presence of graphite, because the PAH molecules
are physically and optically the low-mass end of the graphite particles
\citep{dl01}. Although the exact route of the transformation is unclear,
the anti-spatial correlation of PAH and diamond in Elias~1 implies the
formation of diamonds from carbonaceous structures under the radiation
of high-energy particles.  \citet{ker02} analyzed the precise wavelength
of the 3.53~$\mu$m emission and concluded that the diamonds seen in
HD~97048 are in a hot ambience around 1000~K \citep{lin96}. The
temperature is somehow coincident with the ambient temperature of the
experiment by \citet{ban96} (800~K) and the temperature required for
surfacing diamond via irradiation-induced phase transformation
($>$600~K).


Elias~1 showed a large X-ray flare of $L_X=2.4\times
10^{31}$~erg~s$^{-1}$ with a duration of 10~ks during 120~ks
observations by XMM {\it Newton} in 2001. A similar flare was also
recorded by {\it Chandra} in 2002 \citep[both reported in][]{gia04}. The
duration, duty cycle, peak energy flux, and the plasma temperature are
statistically consistent with the flares seen in solar-type pre-main
sequence stars \citep{fei02,wol05}.
A solar flare accelerates electrons above 20~keV that carry away a total
of $\sim 2 \times 10^{29}$~ergs in a single event while emitting a
similar amount of energy in the soft X-ray regime \citep[$\sim 5 \times
10^{29}$~ergs; ][]{lin74}. About 0.1~\% of the accelerated electrons
reach the interplanetary space \citep{lin71}. If we scale the number of
electrons produced in the solar flare to the flare in Elias~1 referring
to $L_X$,
and with the assumed differential spectral energy distribution with the
power-low index $\sim$3 \citep{lin85,kru07}, the total electron emission
more energetic than 1~MeV comes to $4\times 10^{24}$ erg~s$^{-1}$.
The current influx ($\sim 400$ A~cm$^{-2}$) that the carbon particles
experience during a single event ($\sim$3~hr) are of the same order of
magnitude as those of the experiment by \citet{ban96}
(20--200~A~cm$^{-2}$ for 1~hr) at the distance of 10~AU from Elias~1
where the temperature of disk atmosphere exceeds $~$1000~K \citep[][warm
enough to facilitate the irradiation-induced phase
transformation]{aik06,jon07}. If we count in the protons and the
heavy-ion nuclei, which are 10$^5$ to 10$^6$ times more efficient in
transforming carbon onions to diamonds \citep{wes97}, diamond nucleation
and growth could take place even in the larger area of the
protoplanetary disk.


\acknowledgments We thank all the staff and crew of the Subaru
Telescope for their valuable assistance in obtaining the
data. We thank the anonymous referees for constructive
criticism. Special thanks go to Manuel G\"udel and Sigeo
Yamauchi for enlightening discussions on the X-ray nature of
Herbig Ae/Be stars. We thank the anomymous referees for their
constructive criticism. We appreciate the hospitality of the
Hawaiian local community that made the research presented here
possible. The Dark Cosmology Centre is funded by the Danish
National Research Foundation. M.G. was supported by a fellowship
from the Japan Society for the Promotion of Science.

\clearpage

\end{document}